\RequirePackage{ifpdf}
\ifpdf
\documentclass[cits]{PoS-pdf}
\else
\documentclass[cits]{PoS}
\fi

\usepackage{amsmath,amssymb}
\usepackage{graphicx}

\ifpdf
\usepackage{epstopdf}
\else
\fi

\newcommand{\bq}{\begin{eqnarray}}
\newcommand{\eq}{\end{eqnarray}}
\newcommand{\eps}{\varepsilon}

\title{From motives to differential equations for loop integrals}

\ShortTitle{From motives to differential equations for loop integrals}

\author{S. M\"uller-Stach$^a$, \speaker{S. Weinzierl} $^b$, R. Zayadeh$^a$\\
        \llap{$^a$} PRISMA Cluster of Excellence ,Institut f\"ur Mathematik, Johannes Gutenberg-Universit\"at Mainz\\
        \llap{$^b$} PRISMA Cluster of Excellence, Institut f\"ur Physik, Johannes Gutenberg-Universit\"at Mainz\\
         \email{stach@uni-mainz.de\\
                stefanw@thep.physik.uni-mainz.de\\
                zayadeh@mathematik.uni-mainz.de}}

\abstract{
In this talk we discuss how ideas from the theory of mixed Hodge structures can be used to find differential equations
for Feynman integrals.
In particular we discuss the two-loop sunrise graph in two dimensions and show that these methods lead to a differential equation
which is simpler than the ones obtained from integration-by-parts.
}

\FullConference{"Loops and Legs in Quantum Field Theory" 11th DESY Workshop on Elementary Particle Physics \\
                 April 15-20, 2012\\
                 Wernigerode, Germany}

\begin{document}

\section{Introduction}

The mathematical structures behind Feynman integrals are fascinating and by far not yet completely understood.
Feynman integrals evaluate to transcendental functions and a straightforward question is which
classes of transcendental functions occur in the computation of Feynman integrals.
If we restrict ourselves to one-loop integrals, the answer is known:
We encounter only two transcendental functions, the logarithm and the dilogarithm which are defined by
\bq
\label{def_dilog}
 \mathrm{Li}_1(x) & = & - \ln(1-x) = \sum\limits_{n=1}^\infty \frac{x^n}{n},
 \nonumber \\
 \mathrm{Li}_2(x) & = & \sum\limits_{n=1}^\infty \frac{x^n}{n^2},
\eq
and whose arguments are algebraic functions of the momenta and the masses.
The logarithm and the dilogarithm have generalisations, the most obvious one is given by the polylogarithms defined by
\bq
\label{def_polylog}
 \mathrm{Li}_m(x) & = & \sum\limits_{n=1}^\infty \frac{x^n}{n^m}.
\eq
At the next stage of generalisation one encounters multiple polylogarithms defined by \cite{Goncharov_no_note,Borwein}
\bq
\label{def_multiplepolylog}
 \mathrm{Li}_{m_1,m_2,...,m_k}(x_1,x_2,...,x_k) & = & 
 \sum\limits_{n_1 > n_2 > ... > n_k > 0}^\infty 
 \;\;\;
 \frac{x_1^{n_1}}{n_1^{m_1}} \cdot \frac{x_2^{n_2}}{n_2^{m_2}} \cdot ... \cdot \frac{x_k^{n_k}}{n_k^{m_k}}.
\eq
The class of multiple polylogarithms defined by eq.~(\ref{def_multiplepolylog}) plays an important role in the calculation of Feynman integrals
beyond one-loop.
Indeed, many of the known two-loop amplitudes with massless particles can be expressed in terms of these functions.
The multiple polylogarithms have a rich algebraic structure: There are two Hopf algebras, one with a shuffle multiplication induced from the integral representation and
the other one with a quasi-shuffle multiplication induced from the sum representation.
In addition there are convolution and conjugation operations \cite{Moch:2001zr}.

\begin{figure}
\begin{center}
\includegraphics[viewport=250 650 370 735]{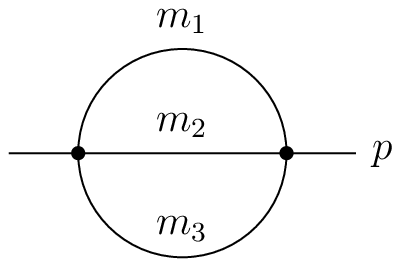}
\end{center}
\caption{
The two-loop sunrise graph.
}
\label{fig_sunrise_graph}
\end{figure}

However, it is also known that there are integrals which cannot be expressed in terms of multiple polylogarithms.
The simplest example is given by the two-loop sunset integral with three internal non-zero masses.
The corresponding Feynman diagram is shown in fig.~(\ref{fig_sunrise_graph}).
It is therefore worth to study this integral in order to learn more about the functions beyond multiple polylogarithms associated to Feynman integrals.
The two-loop sunrise integral has received in the past significant attention 
in the literature \cite{Broadhurst:1993mw,Caffo:1998du,Davydychev:1999ic,Caffo:1999nk,Caffo:2001de,Onishchenko:2002ri,Argeri:2002wz,Caffo:2002ch,Czyz:2002re,Laporta:2004rb,Pozzorini:2005ff,Caffo:2008aw,Groote:2005ay}.
Despite this effort, an analytical answer in the general case of unequal masses is not yet known.
In the special case where all three internal masses are equal, 
a second-order differential equation in the external momentum squared and its analytical solution are known \cite{Laporta:2004rb}.
The analytical solution for the equal mass case involves elliptic functions.
In the general case of unequal masses integration-by-parts identities \cite{Tkachov:1981wb,Chetyrkin:1981qh}
can be used to relate integrals with different powers of the propagators.
In the case of the sunrise topology with unequal masses all integrals can be expressed in terms of four master integrals plus simpler integrals.
This results in a coupled system of four first-order differential equations for the four master integrals \cite{Caffo:1998du}.
This is however not yet the simplest form for the differential equations governing the two-loop sunrise integral.
In a recent publication \cite{MullerStach:2011ru} we showed -- using methods of algebraic geometry -- that also in the unequal mass case 
there is a single second-order differential equation.
In this talk we review the derivation of the second-order differential equation.

Algebraic geometry studies the zero sets of polynomials. A simple example is given by the equation
\bq
\label{simple_example}
 x_1 x_2 + x_2 x_3 + x_3 x_1 & = & 0.
\eq
The zero set of this equation defines an algebraic variety.
In this example it is easily observed that given any solution $(x_1^{(0)},x_2^{(0)},x_3^{(0)})$ also the point 
$(\lambda x_1^{(0)},\lambda x_2^{(0)},\lambda x_3^{(0)})$ is a solution. 
Therefore eq.~(\ref{simple_example}) defines an algebraic variety in the projective space ${\mathbb P}^2$.
We may then study integrals of the form
\bq
\label{simple_integral}
 \int\limits_{x_i \ge 0} d^3x \; \delta\left(1-\sum\limits_{i=1}^3 x_3 \right) \frac{1}{x_1 x_2 + x_2 x_3 + x_3 x_1},
\eq
where the polynomial defining the algebraic variety appears in the denominator.
This leads immediately to the question what happens in the points $(1,0,0)$, $(0,1,0)$ or $(0,0,1)$, which lie on the
boundary of the integration region and where the polynomial in the denominator vanishes.
The interplay between the algebraic variety defined by the zero set of the denominator with the integration boundary 
will be the Leitmotiv in our treatment of the two-loop sunrise integral.

\section{The two-loop sunrise integral}

The two-loop sunrise integral is given in $D$-dimensional Minkowski space by
\bq
\label{def_sunrise}
 S\left( D, p^2 \right)
 & = &
 \left(\mu^2\right)^{3-D}
 \int \frac{d^Dk_1}{i \pi^{\frac{D}{2}}} \frac{d^Dk_2}{i \pi^{\frac{D}{2}}}
 \frac{1}{\left(-k_1^2+m_1^2\right)\left(-k_2^2+m_2^2\right)\left(-\left(p-k_1-k_2\right)^2+m_3^2\right)}.
\eq
Here we suppressed on the l.h.s. the dependence on the internal masses $m_1$, $m_2$ and $m_3$ and on the arbitrary scale $\mu$.
It is convenient to denote the momentum squared by $t=p^2$.
We can trade the integration over the loop momenta for an integration over Feynman parameters.
The integral over the Feynman parameters depends then on two graph polynomials \cite{Bogner:2010kv}
and reads
\bq
\label{def_Feynman_integral}
 S\left( D, t\right)
 & = & 
 \Gamma\left(3-D\right)
 \left(\mu^2\right)^{3-D}
 \int\limits_{\sigma} \frac{{\cal U}^{3-\frac{3}{2}D}}{{\cal F}^{3-D}} \omega,
\eq
where the two Feynman graph polynomials are given by
\bq
 {\cal F} = - x_1 x_2 x_3 t
                + \left( x_1 m_1^2 + x_2 m_2^2 + x_3 m_3^2 \right) {\cal U},
 & &
 {\cal U} = x_1 x_2 + x_2 x_3 + x_3 x_1.
\eq
The differential two-form $\omega$ is given by
\bq
 \omega & = & x_1 dx_2 \wedge dx_3 + x_2 dx_3 \wedge dx_1 + x_3 dx_1 \wedge dx_2.
\eq
The integration is over
\bq
 \sigma & = & \left\{ \left[ x_1 : x_2 : x_3 \right] \in {\mathbb P}^2 | x_i \ge 0, i=1,2,3 \right\}.
\eq
It is simpler to consider this integral first in $D=2$ dimensions and to obtain the result in $D=4-2\eps$ dimensions with the help 
of dimensional recurrence relations \cite{Tarasov:1996br,Tarasov:1997kx}.
In two dimensions this integral is finite and given by
\bq
\label{def_Feynman_integral_dim_two}
 S\left( 2, t\right)
 & = & 
 \mu^2
 \int\limits_{\sigma} \frac{\omega}{{\cal F}}.
\eq
In two dimensions the integral depends only on the second Symanzik polynomial ${\cal F}$.
Note that eq.~(\ref{def_Feynman_integral_dim_two}) is similar to eq.~(\ref{simple_integral}).

Now let us turn to Hodge structures.
Hodge structures have their origin in the study of compact K\"ahler manifolds.
There one has the following decomposition of the cohomology groups
\bq
 H^k\left(X,{\mathbb C}\right) & = & \bigoplus\limits_{p+q=k} H^{p,q}(X),
 \;\;\;\;\;\;
 \overline{H^{p,q}(X)} = H^{q,p}(X).
\eq
For a fixed $k$ this provides an example of a pure Hodge structure of weight $k$.
Algebraic varieties have a generalisation of this structure, which are called mixed Hodge structures \cite{Deligne:1970,Deligne:1971}.
In addition, we can consider a family of Hodge structures, parametrised by a manifold \cite{Griffiths:1968i,Griffiths:1968ii}. 
This is called a variation of a Hodge structure.
Suppose that the cohomology groups are finite dimensional. 
It follows that if a Hodge structure varies smoothly with some parameters, 
then 
\bq
\dim H^{p,q}
\eq
remains constant.
In the following we will relate the order of the differential equation for the two-loop sunrise integral to the dimension of a cohomology group.
If the variation with the internal masses is smooth and if the integral has a second-order differential equation in the equal mass case, it follows
that there must be also a second-order differential equation in the unequal mass case.

Now let us come back to the integral in eq.~(\ref{def_Feynman_integral_dim_two}).
From the point of view of algebraic geometry there are two objects of interest:
On the one hand the domain of integration $\sigma$ and on the other hand the algebraic variety $X$ defined by the zero set of ${\cal F}=0$.
The two objects $X$ and $\sigma$ intersect at the three points
$[1:0:0]$, $[0:1:0]$ and $[0:0:1]$.
\begin{figure}
\begin{minipage}{0.45\textwidth}
\begin{center}
\includegraphics[viewport=230 580 370 740]{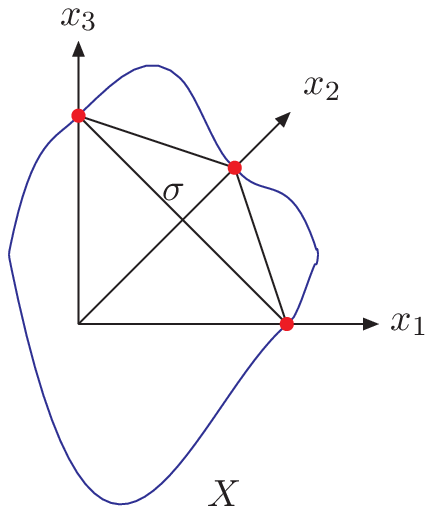}
\caption{
The intersection of the domain of integration $\sigma$ with the zero set $X$ of the second Symanzik polynomial.
}
\label{fig_domain}
\end{center}
\end{minipage}
\hspace*{5mm}
\begin{minipage}{0.45\textwidth}
\begin{center}
\includegraphics[viewport=185 570 345 730]{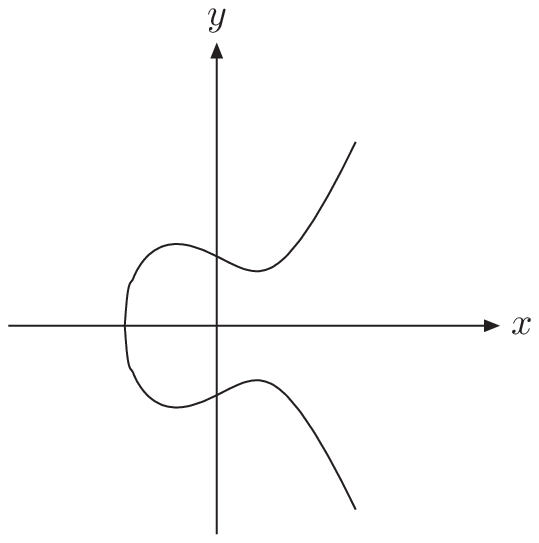}
\caption{
The elliptic curve $y^2=x^3-x+1$.
}
\label{fig_elliptic_curve}
\end{center}
\end{minipage}
\end{figure}
This is shown in fig.~(\ref{fig_domain}).
We blow-up ${\mathbb P}^2$ in these three points and we denote the blow-up by $P$.
We further denote the strict transform of $X$ by $Y$ and the total transform of the set 
$\{ x_1 x_2 x_3 = 0 \}$ by $B$.
With these notations we can now consider the mixed Hodge structure (or the motive) given by the relative cohomology group \cite{Bloch:2006}
\bq
 H^2\left(P \backslash Y, B \backslash B \cap Y \right).
\eq
In the case of the two-loop sunrise integral considered here essential information on 
$H^2(P \backslash Y, B \backslash B \cap Y )$ is already given by $H^1(X)$.
We recall that the algebraic variety $X$ is defined by the second Symanzik polynomial:
\bq
 - x_1 x_2 x_3 t + \left( x_1 m_1^2 + x_2 m_2^2 + x_3 m_3^2 \right) \left( x_1 x_2 + x_2 x_3 + x_3 x_1 \right) & = & 0.
\eq
This defines for generic values of the parameters $t$, $m_1$, $m_2$ and $m_3$ an elliptic curve.
The elliptic curve varies smoothly with the parameters $t$, $m_1$, $m_2$ and $m_3$.
By a birational change of coordinates this equation can brought into the Weierstrass normal form
\bq
 y^2 z - x^3 - a_2(t) x z^2 - a_3(t) z^3 & = & 0.
\eq
The dependence of $a_2$ and $a_3$ on the masses is not written explicitly.
In the chart $z=1$ this reduces to
\bq
\label{weierstrass_normal_form}
 y^2 - x^3 - a_2(t) x - a_3(t) & = & 0.
\eq
The curve varies with the parameter $t$.
An example of an elliptic curve is shown in fig.~(\ref{fig_elliptic_curve}).
It is well-known that in the coordinates of eq.~(\ref{weierstrass_normal_form}) the cohomology group $H^1(X)$ 
is generated by
\bq
 \eta = \frac{dx}{y}
 & \mbox{and} &
 \dot{\eta} = \frac{d}{dt} \eta.
\eq
Since $H^1(X)$ is two-dimensional it follows that $\ddot{\eta}=\frac{d^2}{dt^2} \eta$ must be a linear combination of $\eta$ and $\dot{\eta}$.
In other words we must have a relation of the form
\bq
 p_0(t) \ddot{\eta} + p_1(t) \dot{\eta} + p_2(t) \eta & = & 0.
\eq
The coefficients $p_0(t)$, $p_1(t)$ and $p_2(t)$ define the Picard-Fuchs operator
\bq
 L^{(2)} & = & p_0(t) \frac{d^2}{dt^2} + p_1(t) \frac{d}{dt} + p_2(t).
\eq
Applying the Picard-Fuchs operator to our integrand gives an exact form:
\bq
 L^{(2)} \left(  \frac{\omega}{{\cal F}} \right)
 & = & 
 d \beta.
\eq
The integration over $\sigma$ yields
\bq
 L^{(2)} S(2,t) & = & \mu^2 \int\limits_\sigma d \beta = \mu^2 \int\limits_{\partial \sigma} \beta
\eq
The integration of $\beta$ over $\partial \sigma$ is elementary and we arrive at
\bq
\label{final_result}
 \left[ p_0(t) \frac{d^2}{d t^2} + p_1(t) \frac{d}{dt} + p_2(t)  \right] S\left(2,t\right) & = & \mu^2 p_3(t).
\eq
This is the sought-after second-order differential equation.
The polynomials $p_j(t)$ are given by
\bq
 p_0(t) & = &
  t \left[ t - \left( m_1+m_2+m_3\right)^2 \right]
    \left[ t - \left( -m_1+m_2+m_3\right)^2 \right]
    \left[ t - \left( m_1-m_2+m_3\right)^2 \right]
 \nonumber \\
 & &
    \left[ t - \left( m_1+m_2-m_3\right)^2 \right]
    \left[ 3 t^2 - 2 M_{100} t - M_{200} + 2 M_{110} \right],
 \nonumber \\
 p_1(t) & = & 
  9 t^6
  - 32 M_{100} t^5
  + \left( 37 M_{200} + 70 M_{110} \right) t^4
  - \left( 8 M_{300} + 56 M_{210} + 144 M_{111} \right) t^3
 \nonumber \\
 & &
  - \left( 13 M_{400} - 36 M_{310} + 46 M_{220} - 124 M_{211} \right) t^2
 \nonumber \\
 & &
  - \left( -8 M_{500} + 24 M_{410} - 16 M_{320} - 96 M_{311} + 144 M_{221} \right) t
 \nonumber \\
 & &
  - \left( M_{600} - 6 M_{510} + 15 M_{420} - 20 M_{330} + 18 M_{411} - 12 M_{321} - 6 M_{222} \right),
 \nonumber \\
 p_2(t) & = &
  3 t^5
  - 7 M_{100} t^4
  + \left( 2 M_{200} + 16 M_{110} \right) t^3
  + \left( 6 M_{300} - 14 M_{210} \right) t^2
 \nonumber \\
 & &
  - \left( 5 M_{400} - 8 M_{310} + 6 M_{220} - 8 M_{211} \right) t
  + \left( M_{500} - 3 M_{410} + 2 M_{320} + 8 M_{311} - 10 M_{221} \right),
 \nonumber \\
 p_3(t) & = & 
 -18 t^4
 + 24 M_{100} t^3
 + \left( 4 M_{200} - 40 M_{110} \right) t^2
 + \left( - 8 M_{300} + 8 M_{210} + 48 M_{111} \right) t
 \nonumber \\
 & & 
 + \left( - 2 M_{400} + 8 M_{310} - 12 M_{220} - 8 M_{211} \right)
 + 2 c\left(t,m_1,m_2,m_3\right)  \ln \frac{m_1^2}{\mu^2}
 \nonumber \\
 & &
 + 2 c\left(t,m_2,m_3,m_1\right)  \ln \frac{m_2^2}{\mu^2}
 + 2 c\left(t,m_3,m_1,m_2\right)  \ln \frac{m_3^2}{\mu^2},
\eq
with
\bq
\lefteqn{
c\left(t,m_1,m_2,m_3\right) = } & &
 \nonumber \\
 & &
 \left( -2 m_1^2 + m_2^2 + m_3^2 \right) t^3
 + \left( 6 m_1^4 - 3 m_2^4 - 3 m_3^4 - 7 m_1^2 m_2^2 - 7 m_1^2 m_3^2 + 14 m_2^2 m_3^2 \right) t^2
 \nonumber \\
 & &
 + \left( -6 m_1^6 + 3 m_2^6 + 3 m_3^6 + 11 m_1^4 m_2^2 + 11 m_1^4 m_3^2 - 8 m_1^2 m_2^4 - 8 m_1^2 m_3^4 - 3 m_2^4 m_3^2 - 3 m_2^2 m_3^4 \right) t
 \nonumber \\
 & & 
 + \left( 2 m_1^8 - m_2^8 - m_3^8 - 5 m_1^6 m_2^2 - 5 m_1^6 m_3^2 + m_1^2 m_2^6 + m_1^2 m_3^6 + 4 m_2^6 m_3^2 + 4 m_2^2 m_3^6 
 \right. \nonumber \\
 & & \left.
        + 3 m_1^4 m_2^4 + 3 m_1^4 m_3^4 - 6 m_2^4 m_3^4 
        + 2 m_1^4 m_2^2 m_3^2 - m_1^2 m_2^4 m_3^2 - m_1^2 m_2^2 m_3^4 \right).
\eq
In order to present the result in a compact form we have introduced 
the monomial symmetric polynomials $M_{\lambda_1 \lambda_2 \lambda_3}$ in the variables $m_1^2$, $m_2^2$ and $m_3^2$.
These are defined by
\bq
 M_{\lambda_1 \lambda_2 \lambda_3} & = &
 \sum\limits_{\sigma} \left( m_1^2 \right)^{\sigma\left(\lambda_1\right)} \left( m_2^2 \right)^{\sigma\left(\lambda_2\right)} \left( m_3^2 \right)^{\sigma\left(\lambda_3\right)},
\eq
where the sum is over all distinct permutations of $\left(\lambda_1,\lambda_2,\lambda_3\right)$.
In the equal mass case eq.~(\ref{final_result}) reduces to the well-known result of \cite{Laporta:2004rb}.
The differential equation in eq.~(\ref{final_result}) has been confirmed with numerical methods in \cite{Groote:2012pa}.

\bibliography{/home/stefanw/notes/biblio}
\bibliographystyle{/home/stefanw/latex-style/h-physrev5}

\end{document}